\documentclass[journal,onecolumn]{IEEEtran}
\usepackage{amsmath}
\usepackage{amssymb}
\usepackage{graphicx}
\usepackage{subfigure}
\usepackage{multirow}
\usepackage{setspace}
\usepackage{amsfonts}
\usepackage{cite}
\usepackage{bm}
\usepackage{booktabs}
\usepackage{makecell}

\begin{document}
\title{A Neural Speech Codec for Noise Robust Speech Coding}
\author{Jiayi Huang, Zeyu Yan, Wenbin Jiang, He Wang, 
        Fei Wen%, \textit{Senior Member}, \textit{IEEE}
        
\IEEEcompsocitemizethanks{
\IEEEcompsocthanksitem %Corresponding author: Fei Wen.\\  
%This work was supported in part by the National Natural Science Foundation of China (NSFC) under Grant 62271314. 
%Corresponding author: Fei Wen
J. Huang, Z. Yan, P. Liu, and F. Wen are with the Brain-Inspired Application Technology Center (BATC), School of Electronic Information and Electrical Engineering, Shanghai Jiao Tong University, Shanghai, China. W. Jiang is with Hangzhou Dianzi University, Hangzhou, China. H. Wang is with Shanghai Aero Measurement Controlling Research Institute.
E-mail: sycamore2017@sjtu.edu.cn; zeyuyan@sjtu.edu.cn; jwb361@sjtu.edu.cn;  wenfei@sjtu.edu.cn.
}% <-this % stops an unwanted spac
\thanks{}}
        
% The paper headers
\markboth{}
{Shell \MakeLowercase{\textit{et al.}}: Bare Demo of IEEEtran.cls for Journals}

\maketitle

\begin{abstract}
This paper considers the joint compression and enhancement 
problem for speech signal in the presence of noise.
Recently, the SoundStream codec, which relies on
end-to-end joint training of an encoder-decoder pair and 
a residual vector quantizer
by a combination of adversarial and reconstruction losses,
has shown very promising performance, especially in subjective perception quality.
In this work, we provide a theoretical result to show that, 
to simultaneously achieve low distortion and high perception in the presence of noise, 
there exist an optimal two-stage  optimization procedure 
for the joint compression and enhancement problem. 
This procedure firstly optimizes an encoder-decoder
pair using only distortion loss and then fixes the encoder
to optimize a perceptual decoder using perception loss.
Based on this result, we construct a two-stage training framework 
for joint compression and enhancement of noisy speech signal.
Unlike existing training methods which are heuristic,
the proposed two-stage training method has a theoretical foundation.
Finally, experimental results for various noise and bit-rate conditions 
are provided. The results demonstrate that a codec trained by
the proposed framework can outperform SoundStream and other representative codecs 
in terms of both objective and subjective evaluation metrics.
Code is available at \textit{https://github.com/jscscloris/SEStream}.
%For audio with background noise, the traditional speech codec needs to remove noise through a speech enhancement network previously to obtain clean speech. In this paper, we present an optimal training framework of a joint noise reduction compression, which can reduce the noise during the compression of audio codec. This framework divides training into two stages. The first stage trains an optimal encoder to achieve maximum possible noise reduction. In the second stage, the optimal encoder is fixed, and then a decoder with the best perceptual quality is trained under the optimal perceptual constraint. We prove the optimality of this framework by formula derivation. And experiments show that the proposed framework outperforms Soundstream and other codec in both objective and subjective evaluation metircs.
\end{abstract}
\begin{IEEEkeywords}
Speech compression, speech enhancement, noise reduction, codec, neural networks
\end{IEEEkeywords}

\IEEEpeerreviewmaketitle {}

\section{Introduction}
Audio codec is a fundamental tool for audio transmission. 
Recently, benefited from the powerful deep learning techniques,
neural networks based audio codecs have made much progress and
shown considerable improvement over traditional digital signal 
processing based codecs.
Generally, neural networks based codecs can be divided into two categories.
The first is based on generative models \cite{Lyra,wavenet0,sourcecoding,sampleRNN,gopus,glpcnet},
which aims to generate high-fidelity speech from compressed representation.
The second is based on end-to-end neural network models,
which learns an encoder-decoder pair in an end-to-end manner \cite{vaewavenet,semivae,soundstream,cascadedcoding, Lin2022SpeechEF, Chen2021TeNCLB}.
Generative models, such as WaveNet \cite{wavenet}, WaveRNN \cite{wavernn}, 
and WaveGRU \cite{Lyra}, are typically combined with traditional codec algorithms. 
Given compressed representation from traditional codecs, 
these models have shown impressive performance in generating high perception quality audio \cite{Lyra,wavenet0,sourcecoding,sampleRNN,gopus,glpcnet}.
End-to-end models \cite{vaewavenet,semivae,soundstream,cascadedcoding} 
aim to encode audio into a bit stream at a target bit-rate, and then 
recover the source audio from the bit stream in an end-to-end way using methods like variational autoencoder (VAE) \cite{vqvae1,VQ-VAE}. 
While these methods have achieved remarkable performance, 
how to construct an optimal training framework for end-to-end codec model
in the presence of noise is still an open question.

%Traditionally, codecs typically transmit audio using digital signal processing. However, neural network-based codecs are progressively replacing traditional parametric and waveform codecs due to their excellent ability to extract complex features. These neural network-based codecs can be divided into two categories. The first is based on generative models\cite{Lyra,wavenet0,sourcecoding,sampleRNN,gopus,glpcnet}, and the other is based on end-to-end neural network Generative models like WaveNet\cite{wavenet}, WaveRNN\cite{wavernn}, and WaveGRU\cite{Lyra}, combined with traditional codec algorithms, are widely used in neural codecs\cite{Lyra,wavenet0,sourcecoding,sampleRNN,gopus,glpcnet} to generate high-quality audio.models\cite{vaewavenet,semivae,soundstream,cascadedcoding}. 
%Another solution\cite{vaewavenet,semivae,soundstream,cascadedcoding} calls for encoding the audio as a bit stream at the target bit-rate, then recovering the original audio from the bit stream in an end-to-end way using methods like variational autoencoder (VAE)\cite{vqvae1,VQ-VAE}. While these methods have achieved remarkable results, they still face two main challenges: 1) the trade-off among compression rate, distortion, and perceptual quality; and 2) the background noise in the speech. 

High perception quality is a main goal of audio codecs,
which measures the degree to which the decoded audio sounds like
natural clean audio from human subjective judgement.
Taking perception quality into consideration,
audio codecs typically aim to encode audio into as few bits 
as possible and, at meantime, reconstruct audio with less 
distortion and better perceptual quality as possible.
Intuitively, the quality of the reconstructed audio depends on the bit-rate. 
The lower the  bit-rate, the worse the quality of the reconstructed audio. 
Recently, it has been shown both theoretically and empirically that,
at a given bit-rate, there is a tradeoff between distortion and perception \cite{pdtradeoff, blau2019rethinking, yan2021perceptual,yan2022optimally}. 
That is, to achieve high perception quality, an elevation of the lowest achievable 
distortion is necessary.
Accordingly, there is three way tradeoff between bit-rate, distortion,
and perceptual quality.
To achieve high perception quality,
the recent work SoundStream \cite{soundstream} 
employs a combination of adversarial and reconstruction losses
to train an end-to-end encoder-decoder model, 
in which a residual vector quantizer is jointly trained  
for bit-rate scalable codec.
SoundStream has shown remarkable performance in perception quality especially in 
low bit-rate case.

%According to the research on lossy compression, there is also a trade-off between distortion and perceptual quality. The sacrifice of distortion is inevitable in order to obtain higher perceptual quality.
%The trade-off problem exists widely across various codecs. Audio codecs aim to encode audio in as few bits as possible and reconstruct audio with less distortion and better perceptual quality. Usually, the quality of the reconstructed audio is proportional to the bit-rate. The lower the target bit-rate, the worse the quality of the reconstructed audio. According to the research on lossy compression, there is also a trade-off between distortion and perceptual quality. The sacrifice of distortion is inevitable in order to obtain higher perceptual quality. A mutual constraint exists among compression rate, distortion, and perceptual quality. To deal with these constraints, some generative algorithms \cite{soundstream} uses adversarial loss to represent perceptual quality and then balance distortion and perception by adjusting the weights of spectral reconstruction loss and adversarial loss. However, parameter tuning is frequently expensive and difficult to approach the optimal upper bound.

In many practical applications, audio codecs inevitably suffer from 
background noise from the environment. In the presence of noise, 
a straightforward approach is to use a cascade of a denoising model and a codec, which 
firstly employs a denoising model to suppress the noise in the noisy signal and, 
then, processes the denoised signal by the codec. 
In theory, this combination approach using two cascaded models is optimal in compressing 
noisy signal \cite{dobrushin1962information,wolf1970transmission}.
However, in practice, using two models for denoising and compression respectively is more complex and would incur larger latency compared with a single model for joint denoising and compression.
Recently, some works have studied the joint speech enhancement and compression problem using neural models. 
For example, based on VQ-VAE autoencoder with WaveRNN decoder,
the work \cite{enhancingcodec} 
proposes a compressor-enhancer encoder for audio codec in noisy conditions.
The work \cite{jiang2022end} 
proposes an end-to-end neural speech codec with
low latency, namely TFNet,  
which is jointly optimized with speech enhancement and
packet loss concealment.
In \cite{omran2022disentangling}, 
the separation of speech signal from background noise 
in compressed domain of a neural audio codec has been considered.
Moreover, SoundStream \cite{soundstream} shows that compression and enhancement can be jointly considered, which achieves controllable noise reduction through a FiLM layer \cite{film}. 

Though straightforward training of an end-to-end codec model using noisy speech 
data can naturally make the learned model robust to noise, 
how to construct an optimal end-to-end training framework 
for joint speech compression and enhancement in the presence of noise 
is still an open question.
Existing methods typically use a heuristic combination of distortion and adversarial losses to achieve high perception quality, which lacks theoretical foundation.

This paper provides a theoretical
analysis on the joint speech compression and enhancement problem,
which considers a lossy compression model under both distortion and 
perception quality constraints. The analysis sheds some light on
how to construct an optimal training framework for speech compression
in the presence of noise. Based on the result,
we develop a two-stage training framework and evaluate it
in various bit-rate and noise conditions in comparison with existing methods.

The main contributions of this work are as follows.
\begin{itemize}

\item
We provide a theory for optimally training of joint speech compression and enhancement in the presence of noise.
The theoretical result reveals that to simultaneously achieve 
low distortion and high perception quality
for joint compression and enhancement in the presence of noise, 
an optimal optimization procedure is given by a two-stage optimization
that, firstly, optimizes an encoder-decoder pair using only distortion loss 
and, then, fixes the obtained encoder to optimize a perceptual decoder using perception loss.

\item
Based on the theoretical result, we construct a two-stage training 
framework for joint compression and enhancement of noisy speech signal.
To achieve satisfactory performance on speech signal, 
multi-scale time-frequency spectrum based reconstruction loss and multi-scale adversarial loss are employed.

\item
The proposed method is evaluated in comparison with 
state-of-the-art speech codecs, including SoundStream,
Lyra, EVS, and OPUS.
Experimental results under various noise and bit-rate conditions  
show that the proposed two-stage method can achieve better 
performance in terms of both objective and subjective 
evaluation metrics than state-of-the-arts.

%\item We propose an optimal training framework to integrate enhancement and compression to achieve the upper bound under optimal perception, and prove it theoretically.

%\item We extend the theory of lossy compression and verify it under free noisy conditions.

%\item We compare the proposed model with other advanced audio codecs, and our method has advantages in both objective and subjective merics.
\end{itemize}

\section{Related work}

\subsection{Audio Codecs}
Audio codec is a fundamental technique in audio communication, 
which aims to compress audio signal at a given bit-rate with
distortion as less as possible.
Classic audio codecs include USAC\cite{USAC}, EVS\cite{EVS} and OPUS\cite{OPUS}. USAC (Unified Speech and Audio Coding) is a low-delay, high-quality audio coding standard developed by MPEG.
OPUS is an open-source audio codec with sampling rates ranging from 6 kbps to 510 kbps. It supports a wide range of audio applications from video conferencing to streaming media services. The performance of OPUS is excellent in high bit-rate compression.
EVS (Enhanced Voice Services) is the latest codec developed by 3GPP standardization organization for mobile networks. It supports target bit-rates ranging from 5.9 kbps to 128 kbps and performs better than OPUS at medium and low target bit-rate. 

Generative model has become a popular tool for audio reconstruction 
tasks due to its ability to generate high-quality audios. 
Wavenet\cite{wavenet,parallelwavenet} is an auto-regressive 
sequence generation model, which can directly generate high-quality 
speech waveforms from input signal. Based on it, a series of 
generative models such as WaveRNN\cite{wavernn}, WaveGAN\cite{wavegan}, 
and Parallel WaveGAN\cite{parallelwavegan} have been developed as 
vocoder for text-to-speech, which further improve the speed of 
processing and quality of the reconstructed audio. 
Lyra\cite{Lyra} utilizes an auto-regressive WaveGRU model as 
the decoder of the codec. It can achieve excellent performance 
at a low target bit-rate of 3 kbps. 
Another type of MelGAN-based generative models mainly 
focuses on improved discriminators
for audio signal, which can improve perceptual quality by a large margin.
For example, MelGAN \cite{melgan} proposes to use multi-scale 
discriminators to achieve better perception quality.
%Following that, HiFi-GAN \cite{kong2020hifi} and Fre-GAN \cite{fregan} 
%further optimize the way to extract audio features, 
%such that they can generate audio close to natural speech. 
Furthermore, HIFI-GAN \cite{kong2020hifi} designs a multi-period 
discriminator and proposes a multi-receptive field fusion method. 
Based on HIFI-GAN, Fre-GAN \cite{fregan} uses DWT to retain 
high-frequency information and uses RCG to capture information 
of different frequency bands. These methods have shown effectiveness 
in improving naturalness of generated speech.

Recently, end-to-end neural audio codecs have shown promising performance. 
The work \cite{vaewavenet} proposes a neural network architecture based on 
VQ-VAE and WaveNet decoder. It can generate high-quality audio at a very 
low bit-rate of 1.6 kbps. The work \cite{semivae} introduces a speaker 
encoder and speaker VQ codebook into the VQ-VAE architecture to generalize 
to unseen speakers or content. SoundStream \cite{soundstream} proposes a 
bit-rate scalable codec with VQ-VAE-2 \cite{VQ-VAE}, which can support 
scalable bit-rates from 3 kbps to 18 kbps with a single model by simply 
controlling the number of cascaded VQs during inference. 

\subsection{Speech Enhancement}

Speech enhancement, also known as noise reduction, 
aims to remove the background noise of observed noisy speech. 
It is widely used as a front-end processing technology 
in speech recognition, hearing aids, and telephone communication. 
Generally speaking, speech enhancement methods can be divided 
into two categories, the traditional statistical model based methods \cite{Boll1979SuppressionOA,Berouti1979EnhancementOS,Ephraim1993ASS}, 
and the recent DNN-based methods \cite{denoising0,denoising1,segan,wavenetdenising,TasNet,metricenhance,dnnenhance,metricgan,maskgan,diffusionenhance,Jiang2022EfficientSE,Jiang2022SpeechEW,Donahue2017ExploringSE,reverberant1,reverberant2,superresolution,ganenhancement,bandextension}. 
Traditional methods, such as the spectral subtraction method, 
the statistical model based method, and the subspace based method, 
perform well on the suppression of stationary noise. 
However, these methods often perform poorly on non-stationary noise. 

Benefited from the powerful deep learning technique, 
DNN-based methods can achieve significantly 
better performance than the traditional statistical model based methods \cite{denoising1,segan,wavenetdenising,TasNet,metricenhance,dnnenhance}. 
However, the most popular speech codecs, such as EVS and OPUS,
do not consider speech denoising or use cascaded denoising module. 
OPUS applies discontinuous transmission (DTX) to reduce the impact 
of noise by reducing the bit-rate of silent frames or noisy frames 
and applies noise shaping filters to reduce noise. 
EVS uses DTX and Comfort Noise Generation (CNG) to handle noisy speech. 
It employs Voice Activity Detection (VAD) to detect silent segments or 
background noise, encodes the noise with dedicated cores and finally 
compensates the distortion in speech with CNG. Another popular speech 
codec Lyra still applies an additional module TasNet \cite{TasNet} 
for noise suppression before the encoding the signal. To improve this 
cascaded structure, SoundStream proposes that compression and enhancement 
can be jointly performed by the single model without increasing the overall delay. 
It uses Feature-wise Linear Modulation (FiLM) \cite{film} to achieve 
controllable denoising. The FiLM layer performs a simple feature-wise 
affine transformation on the features of the intermediate layer of the 
neural network and dynamically activates and deactivates the denoising 
with an additional conditioning signal.

\subsection{The Rate-Distortion-Perception Tradeoff}

In speech codec, the objective is to achieve lower distortion and higher perceptual quality at a given bit-rate. 
Recent studies show that distortion and perception quality are at odd with each other. More specifically,
minimizing distortion only would not necessarily lead to good perception quality. Imposing perfect perception quality constraint would lead to increase of the lowest achievable distortion \cite{pdtradeoff,blau2019rethinking}. 
That is high perception quality can only be achieved at the cost of increased distortion.
%However, bit-rate, distortion, and perception are at odd with each other. Recent studies have shown that distortion is not proportional to human perception\cite{pdtradeoff,blau2019rethinking}. 
In the field of audio processing, the commonly used spectral reconstruction loss is also a kind of distortion loss. Therefore, only considering the spectral distortion of the reconstruction is not enough for achieving high perception quality.
%deviation on the spectrum will result in a serious mechanical sense in the audio, which is far from the natural pronunciation of human beings. 
Mathematically, perceptual quality can be expressed as the deviation between the distribution of the decoded signal \(\hat{X}\) and that of the source signal \(X\), as \cite{pdtradeoff}
\begin{equation}
\begin{aligned}
d(p_{X},p_{\hat{X}}),
 \label{eq0} 
\end{aligned}
\end{equation}
where \(d(\cdot,\cdot)\) is a divergence measures the
deviation between two distributions, such as the KL divergence or Wasserstein distance,
which satisfies \(d(p,q)=0\) if and only if \(p=q\).
When \(d(p_{X},p_{\hat{X}} )=0\), \(\hat{X}\) and \(X\)
have the same distribution as \(p_{X}=p_{\hat{X}}\).
In this case, the reconstruction has perfect perception quality.

As generative adversarial training is effective in
aligning distributions,
adversarial loss is widely used in learning deep neural
networks based codecs to minimize the deviation between
the distributions of the decoded output $\hat{X}$
and the source ${X}$, i.e., \(d(p_{X},p_{\hat{X}} )=0\).
It has been shown that, such adversarial training
can significantly improve the naturalness of the generated audio.
However, using only adversarial loss can result in
excessive increase of the reconstruction distortion,
e.g, though the decoded output can have good perception
quality but with large distortion from the input signal.
Therefore, adversarial loss is usually used in combination with
distortion loss such as spectral reconstruction loss,
with a balance between these two losses to achieve a trade-off
between distortion and perception \cite{melgan,kong2020hifi,soundstream}

Recently, the works \cite{yan2021perceptual,yan2022optimally} have studied
the training framework for perceptual lossy compression under perception constraint.
A method for flexibly controllable perception-distortion trade-off  
has been proposed in \cite{yan2022optimally}.
For audio codec, there also exist a perception-distortion trade-off problem,
as empirically shown in \cite{soundstream}.
As the work \cite{yan2021perceptual} considers the lossy compression problem
in the absence of noise, the results do not apply to audio codec in processing
noisy speech signal. Speech noise is inevitable in many real-world applications.
In this work, we extend the work \cite{yan2021perceptual} to consider the
joint compression and enhancement of noisy speech signal. 

\section{Theory and Methodology}

\subsection{Analysis on Joint Compression and Enhancement}

We first review existing results on the compression problem in the 
absence of noise and, then, present extended results on the joint compression and enhancement problem in the presence of noise.

Suppose that \(X\) is a deterministic discrete source to be compressed. In the speech compression task without considering noise, an encoder \(E\) encodes the source signal \(X\) into a sequential compressed representation \(Z\) at a given bit-rate \(R\), and a decoder \(G\) decodes the compressed representation \(Z\) to obtain \(\hat{X}\), which can be written as 
\[X \stackrel{E}{\longrightarrow} Z \stackrel{G}{\longrightarrow} \hat{X}.\] 
At a given bit-rate $R$, the rate-distortion-perception (R-D-P) function for perceptual lossy compression, which additionally takes perception quality constraint into consideration in the Shannon's rate-distortion function, can be expressed as
\begin{equation}
\begin{aligned}
R^{(I)} (D,P)&=\min_{p_{\hat{X}|X}} (X;\hat{X})\\
\mathrm{subject~ to} ~~~~&\mathbb{E} [\Delta(X,\hat{X})]\le D,\\ ~~~~&d(p_{X},p_{\hat{X}} )\le P,
 \label{eq1} 
\end{aligned}
\end{equation}
where \(D\) stands for the distortion constraint, \(P\) stands for the perception quality constraint based on the divergence between $\hat{X}$ and $X$, and \(\Delta\) is a distortion measure function. For the R-D-P problem (\ref{eq1}),
the following result has been derived in \cite{yan2021perceptual}.

%\newtheorem{theorem}{Theorem}
%\begin{theorem}
\textit{Theorem 1 \cite{yan2021perceptual}:}
Suppose that \(X\) is a deterministic discrete source, and
the distortion function \(\Delta\) measures the squared error. 
Let \((E_d,G_d)\) be an optimal encoder-decoder pair to problem (\ref{eq1}) in the case
without any constraint on perception, i.e. \(P=+\infty \) in (\ref{eq1}), 
then \(E_d\) is also an optimal encoder in the case perfect perception constraint i.e. \(P=0\) in (\ref{eq1}). Furthermore, denote $X_{mse}:=G_d(E_d(X))$, there holds the following equations: 
%\textit{Proof:} Let \(X_{mse}=\mathbb{E}[X|\hat{X}]\) be the results of optimal encoder in the case of MSE loss. The distortion constraint can be writtten as:
%\begin{equation}
%\begin{aligned}
%&\mathbb{E}[\Delta(X,\hat{X})]\\
%&=\mathbb{E}\left[\left \|(X-X_{mse} )+(X_{mse}-\hat{X})\right %\| ^{2}\right ]\\
%&={\textstyle \sum_{x,x_{mse}}} p_{X,X_{mse}}(x,x_{mse})\left \|x-x_{mse}\right \|^2\\
%&~~~+{\textstyle \sum_{x_{mse},\hat{x}}} p_{X_{mse},\hat{X}}(x_{mse},\hat{x})\left \|x_{mse}-\hat{x} \right \|^2.
% \label{eq2} 
%\end{aligned}
%\end{equation}
%The first term is the MSE between the source signal and the decoding results of minimum MSE loss . The second term is the MSE loss generated by mapping \(X_{mse}\) to \(\hat{X}\), which is distributed the same as the source. 
\begin{equation}
\begin{aligned}
    p_{X,X_{mse}}=p_{\hat{X},X_{mse}},\\
    p_{X|X_{mse}}=p_{\hat{X}|X_{mse}}.
    \label{eq3}
\end{aligned}
\end{equation}
%\end{theorem}

This result implies that any optimal encoder achieving the minimum mean squared error (MSE) in the case without perception constraint is also an optimal encoder in the case with perfect perception constraint.
That is any optimal encoder to \(R^{(I)}(D,+\infty)\) is also an optimal encoder to \(R^{(I)}(D,0)\).
Next we consider the condition in the presence of noise and provide extended result.
%This theorem has been proved in \cite{yan2021perceptual} and successfully applied in image compression. We can conclude that the optimal encoder achieving the minimum MSE in the case without perception constraint is also an optimal encoder in the case with perfect perceptual quality. Based on this theorem, we extend it as followed.

%\newtheorem{corollary}{Corollary}
%\newtheorem{theorem}{Theorem}
%\begin{theorem}
\textit{Theorem 2:}
Let \(X\) be a deterministic discrete source,
and \(X'\) be a noisy observation from \(X\) as \(X'=X+N\), 
where \(N\) is noise.
Suppose that \(N\) is independent with \(X\) and the distortion function \(\Delta\) measures the squared error, then, any optimal encoder under the condition without any perception constraint, i.e.,
\(P=+\infty\) in (\ref{eq1}), is also an optimal encoder under the condition with perfect perception constraint, i.e., \(P=0\) in (\ref{eq1}).
%\end{theorem}

\textit{Proof:} The minimum MSE codec process in the presence of noise can be expressed as
\[X\longrightarrow X' \stackrel{E}{\longrightarrow} Z \stackrel{D}{\longrightarrow} X_{mse}.\]
Denote the output of the optimal noise reduction by 
\[X_{deN}:=\mathbb{E}[X|X'].\] 
Under the condition that \(\Delta\) measures the squared error, the optimal process for joint compression and noise reduction can be written as
\[X\longrightarrow X_{deN}\longrightarrow X_{mse},\] 
where \(X_{mse}\) is the minimum MSE codec output from \(X_{deN}\).
Given \(X_{deN}\), \(X\) and \(X_{mse}\) are independent of each other since the data process \(X\longrightarrow X_{deN}\longrightarrow X_{mse}\) is a Markov chain.
%according to the principle of Markov chain. 
Then for the distortion term, we have
\begin{equation}
\begin{split}
    &\mathbb{E}[\Delta(X,X_{mse})]\\
    &=\mathbb{E} \left[\left\|(X-X_{deN})+(X_{deN}-X_{mse})\right\|^2\right]\\
    &=\mathbb{E} \left[\left\|(X-X_{deN})\right\|^2\right]+\mathbb{E} \left[\left\|(X_{deN}-X_{mse})\right\|^2\right]\\
    &~~~-2\mathbb{E}_{deN}\Big\langle\mathbb{E}[(X-X_{deN})|X_{deN}], \\
    &~~~~~~~~~~~~~~~~\mathbb{E}[(X_{deN}-X_{mse})|X_{deN}]\Big\rangle. 
    \label{eq4}
\end{split}
\end{equation}
Under the condition that
the noise \(N\) is independent of the clean source \(X\),
we have $\mathbb{E}[X-X_{deN}]=0$ and then it follows from (\ref{eq4}) that
%Assuming that the noise \(N\) is independent of the clean speech \(X\), we suppose that \(X_{deN}\) achieves perfect noise reduction, \(\mathbb{E}[X-X_{deN}]=0\). Then the third term of (\ref{eq4}) equals to 0. Equation (\ref{eq4}) can be simplified as:
\begin{equation}
\begin{aligned}
    &\mathbb{E}[\Delta(X,X_{mse})]\\
    &=\mathbb{E} \left[\left\|(X-X_{deN})\right\|^2\right]
    +\mathbb{E} \left[\left\|(X_{deN}-X_{mse})\right\|^2\right].
    \label{eq5}
\end{aligned}
\end{equation}
It implies that the an end-to-end optimization from \(X'\) to \(\hat{X'}\)
is equivalent to the two-stage optimization process which firstly conduct noise reduction to obtain \(X_{deN}\) and then compresses \(X_{deN}\) by the codec. Thus, we can conclude that under the assumption of independent noise, Theorem 2  holds, and the denoised minimum MSE encoder is still an optimal encoder under the condition with perfect perception constraint.

In speech codec model training,
spectral reconstruction loss is usually used
rather than the MSE loss for distortion measurement.
Particularly, the multi-scale spectral reconstruction 
loss has been widely used in high-fidelity audio synthesis and has show superior performance.
The multi-scale spectral reconstruction loss is 
typically defined as \cite{spectralloss}
%MSE loss is not usually not used as a distortion constraint in audio compression. Taking multi-scale spectral reconstruction loss described in  as an example:
\begin{equation}
\begin{aligned}
    L_{dis}(X,\hat{X}) =&\sum_{s\in 2^6,...,2^{11}}\sum_{t}\left\|X_t^s-\hat{X}_t^s\right\|_1\\
    &+\alpha_s\sum_{t}\left\|\log (X_t^s)-\log(\hat{X}_t^s) \right\|_2,  
    \label{eq6}
\end{aligned}
\end{equation}
where \(X_t^s\) denotes the \(t\)-th frame of the time-frequency spectrum (or Mel-spectrogram) of the input signal \(X\) with a window length equals to \(s\). \(\hat{X}^s\) denotes the spectrogram of the reconstructed signal \(\hat{X}\). \(\alpha_s\) is a balance parameter which is typically chosen to be \(\alpha_s=\sqrt{s/2}\) \cite{soundstream}. The loss (\ref{eq6}) computes the $\ell_1$ distortion of the spectrum and the $\ell_2$ distortion of the log-scale spectrum for multiple scales in the spectral domain.
Though Theorem 2 is only derived for the case of the MSE loss, it still sheds some light on the cases with other distortion losses such as (\ref{eq6}). 

The above analysis considers the MSE loss.
For speech signal processing, 
the multi-scale spectral reconstruction
loss (\ref{eq6}), which consists of $\ell_1$ distance on the spectrogram slice and $\ell_2$ distance on the log of the spectrogram slice, has shown superiority and hence be widely used in practice.
Next, we provide further analysis for the multi-scale spectral reconstruction
loss to show that Theorem 1 in the MSE distortion case can be extended to $\ell_1$ norm case and logarithmic MSE case.

Let \(\hat{X}\) be the decoding output which minimizes a generalized distortion such as the $\ell_1$ distance with respect to the spectrogram slice or the $\ell_2$ distance with respect to the log of the spectrogram slice. Meanwhile, let \(\hat{X}_{p}\) denote the output with perfect perception constraint. According to the principle of Markov chain, \(X\) and \(\hat{X}_{p}\) are independent of each other.

For the case of the $\ell_1$ loss, the distortion is $\ell_1$ distance with respect to the spectrogram slice as
\begin{equation}
\begin{aligned}
\mathbb{E} [\Delta(X,\hat{X}_{p})]&=\mathbb{E}\left[\left\|X-\hat{X}_{p}\right\|\right]\\
&\le\mathbb{E}\left[\left\|X-\hat{X}\right\|\right]+\mathbb{E}\left[\left\|\hat{X}_{p}-\hat{X}\right\|\right].
    \label{eq7}
\end{aligned}
\end{equation}
For the case of the log MSE loss, the distortion is $\ell_2$ distance with respect to the log of the spectrogram slice as
\begin{equation}
\begin{aligned}
&\mathbb{E} [\Delta(X,\hat{X}_{p})]\\
&=\mathbb{E}\left[\left\|\log(X)-\log(\hat{X}_{p})\right\|^2\right]\\
&=\mathbb{E}\left[\left\|\log(X)-\log(\hat{X})\right\|^2\right]+\mathbb{E}\left[\left\|\log(\hat{X}_{p})-\log(\hat{X})\right\|^2\right]\\
&~~~+2\mathbb{E}_{\hat{X}}\Big\langle\mathbb{E}\left[(\log(X)-\log(\hat{X}))|\hat{X}\right],\\
&~~~~~~~~~~~~~~~~\mathbb{E}\left[(\log(\hat{X})-\log(\hat{X}_{p}))|\hat{X}\right]\Big\rangle. 
    \label{eq8}
\end{aligned}
\end{equation}
Under the perfect perceptual constraint, i.e. \(d(p_{X},p_{\hat{X}_{p}})\le 0\), we can derive perfect perceptual quality by  \(p_{\hat{X}_{p}|\hat{X}}=p_{X|\hat{X}}\). In the final expansion of (\ref{eq8}), the angle between the two inner products of the third term is 180°, therefore the third term is negative. Thus it follows from (\ref{eq8}) that
\begin{equation}
\begin{aligned}
\mathbb{E} [\Delta(X,\hat{X}_{p})]\le &\mathbb{E}\left[\left\|\log(X)-\log(\hat{X})\right\|^2\right]\\
&+\mathbb{E}\left[\left\|\log(\hat{X}_{p})-\log(\hat{X})\right\|^2\right].
    \label{eq9}
\end{aligned}
\end{equation}

In (\ref{eq7}) and (\ref{eq9}), the first term is the distortion between the source and the minimum distortion decoding output, and the second term is the distortion caused by mapping \(\hat{X}\) to \(\hat{X}_{p}\) with the same distribution as the source. According to Theorem 1, optimizing the encoder \(E_d\) in the case of \(P=+\infty\) results in minimizing the upper bound of the minimum distortion in the case of \(P=0\). Therefore, for the multi-scale spectral reconstruction loss \(L_{dis}\) in (\ref{eq6}), which is composed of $\ell_1$ loss and log MSE, the minimum distortion encoder \(E_d\) in the case without any perception constraint (i.e. \(P=+\infty\)), can also be used as an approximately optimal encoder in the case with perfect perception constraint (i.e. \(P=0\)).

\subsection{Proposed Training Framework for Joint Speech Compression and Enhancement}

From the above analysis,
in this subsection we propose a two-stage training framework for 
joint speech compression and enhancement based on generative adversarial training, in which a generator consists of an encoder-decoder pair
is the codec model to be learned, as shown in Fig. 1. 

\begin{figure*}
    \centering
    \includegraphics[width=0.75\textwidth]{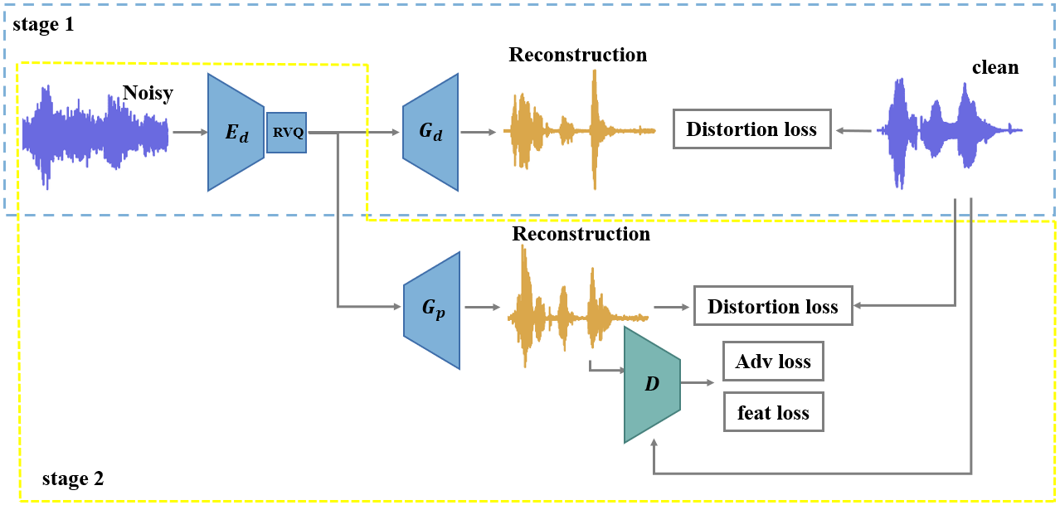}
    \caption{The proposed training framework of SEStream. In the first stage, \((E_d,G_d)\) is trained with distortion loss. In the second stage, \(E_d\) is fixed and $G_p$ is trained with adversarial loss, feature loss and distortion loss jointly. }
    \label{fig:framework}
\end{figure*}

%We chose the framework of SoundStream\cite{soundstream} as the backbone of SEStream. It consists of a generator and two kinds of discriminators. For a fair comparison, the hyper parameters for each module of the model are the same as for SoundStream.

The generator includes an encoder, a residual vector quantizer and a decoder.
It firstly compresses the source signal into compressed representation,
then encodes it with a series of cascaded vector quantizers,
and finally reconstructs the source signal with the decoder.
The encoder consists of a one-dimensional convolution layer
and four encoding modules with stride $(2, 4, 5, 8)$.
Each encoding module consists of three residual units and
a down-sampling layer. Finally, a one-dimensional convolution layer
is used to obtain features with a dimension of 256.
In order to enable real-time inference, all convolutions are causal.
The decoder adopts a similar up-sampling structure.
The residual vector quantizer consists of \(N_q\) cascaded
vector quantizers as in \cite{soundstream}. The target bit-rate
\(R\) of the codec can be calculated by \(R=N_q \cdot S\cdot \log_2(N)\),
where \(N\) is the size of codebook and \(S\) is the number
of frames per second. We also use quantizer dropout in training
to enable bit-rate scalability as in SoundStream \cite{soundstream}.

The discriminator is employed for adversarial training
to improve the perception quality of reconstructed audio.
It consists of a waveform-based discriminator and an
STFT-based discriminator. The waveform-based discriminator
inputs the original audio waveform, two times down-sampled
waveform and four times down-sampled waveform, respectively,
extracting waveform features from different scales. The STFT-based
discriminator first performs STFT transform on the waveform,
and then performs 2-D convolution on the time-frequency spectrum
to extract features.

%The generator includes an encoder, residual vector quantizer and decoder. It first encodes the source signal, then encodes it with a series of cascaded vector quantizers, and finally reconstructs the source signal with a decoder. The encoder consists of a one-dimensional convolution layer and four encoding modules with stride (2, 4, 5, 8). Each encoding module consists of three residual units and a down-sampling layer. Finally, a one-dimensional convolution layer is used to obtain features with a dimension of 256. In order to ensure real-time inference, all convolutions are causal. The decoder adopts a similar structure. The residual vector quantizer consists of \(N_q\) cascaded vector quantizers as described in \cite{soundstream}. The target bit-rate \(R\) of the codec can be calculated by \(R=N_q \times Slog_2N\), where \(N\) is the size of codebook and \(S\) is the number of frames per second. We also apply the quantizer dropout in training to enable bit-rate scalability as used in SoundStream.

%The discriminator is applied to improve the perceptual quality of reconstructed audio. It consists of a waveform-based discriminator and an STFT-based discriminator. The waveform-based discriminator inputs the original waveform, two times down-sampled waveform and four times down-sampled waveform respectively, extracting waveform features from different scales. The STFT-based discriminator first performs STFT transform on the waveform, and then performs 2-D convolution on the time-frequency spectrum to extract features.

Based on the Theorem 2, we propose a two-stage training approach for  end-to-end speech codec learning to achieve joint compression and enhancement, as shown in Fig. 1. The generator is an encoder-decoder pair composed of an encoder \(E\), a quantizer and a decoder \(G\). 
From Theorem 2, an optimal training procedure contains two steps: 

\hangafter 1
\hangindent 1.5em
\textit{i)} In the first step, the encoder-decoder pair \((E_d,G_d)\) is trained  using distortion loss only, e.g., with the multi-scale spectral reconstruction loss \(L_{dis}(\hat{X},X)\) as (\ref{eq6}), without using any discriminator. 

\hangafter 1
\hangindent 1.5em
\textit{ii)} In the second step, the encoder \(E_d\) and the vector quantizer learned in the first stage are frozen, and a new decoder  \(G_p\) is trained by a combination of distortion loss and adversarial loss. 

Though in theory the perceptual decoder can be trained only by adversarial loss, intensive experiments show that a combination of distortion loss and adversarial loss can yield better performance. This is because it is difficult to train a decoder to reconstruct the signal from compressed representation only using adversarial loss.
In the second stage, the overall loss for the generator consists of three components
\begin{equation}
\begin{aligned}
L_G=\lambda_{adv}\cdot L_{adv}(\hat{X},X)&+\lambda_{feat}\cdot L_{feat}(\hat{X},X)\\
&+\lambda_{dis}\cdot L_{dis}(\hat{X},X),
    \label{eq10}
\end{aligned}
\end{equation}
where \(L_{dis}\) is the multi-scale spectral reconstruction loss (\ref{eq6}).
Similar to \cite{soundstream}, we consider 4 individual discriminators,
$\mathcal{D}_{k}$ for $k \in \{0, \cdots, 3\}$,
with the index $k=0$ denoting the STFT-based discriminator and
$k \in \{1, \cdots, 3\}$ denoting waveform-based discriminator for different resolutions.
With these discriminators, \(L_{adv}\) is an adversarial loss 
for the generator as 
\begin{equation}
\begin{aligned}
L_{adv}(\hat{X},X)=\mathbb{E}\left[\frac{1}{K}\sum_{k,t}\frac{1}{T_k}\max(0,1-\mathcal{D}_{k,t}(\hat{X}))\right],
    \label{eq11}
\end{aligned}
\end{equation}
where $T_k$ is the number of logits at the output of the $k$-th discriminator along the time dimension. 

\(L_{feat}\) is a feature loss computed based on the discrepancy between the internal layer outputs of the discriminators, i.e., the difference on the features of the discriminators, as
\begin{equation}
\begin{aligned}
L_{feat}(\hat{X},X)=\mathbb{E}\left[\frac{1}{KL}\sum_{k,l}\frac{1}{T_{k,l}}\sum_{t}\left|\mathcal{D}_{k,l}^{(l)}(X)-\mathcal{D}_{k,l}^{(l)}(\hat{X})\right|\right], 
    \label{eq12}
\end{aligned}
\end{equation}
where $K$ is the number of discriminators, 
$L$ is the number of internal layers of each discriminator.  
%\(\mathcal{D}\) is the discriminator, including three waveform discriminators and an STFT discriminator. 
\(\mathcal{D}_{k,l}\) is the output of \(l\)-th 
layer of the \(k\)-th discriminator. 
The discriminator is trained by a loss as
\begin{equation}
\begin{aligned}
L_D=&\mathbb{E}\left[\frac{1}{K}\sum_{k}\frac{1}{T_k}\sum_{t}\max(0,1-\mathcal{D}_{k,t}(X))\right]\\
&+ \mathbb{E}\left[\frac{1}{K}\sum_{k}\frac{1}{T_k}\sum_{t}\max(0,1+\mathcal{D}_{k,t}(\hat{X}))\right].
    \label{eq13}
\end{aligned}
\end{equation}

In the first stage of training, we aim to minimize the distortion 
by the multi-scale spectral reconstruction loss and achieve  
maximal noise suppression in the encoder \(E_d\). In the second 
stage of training, we use generative adversarial training so that 
the decoder can generate high perception quality audio. In order to 
avoid excessive distortion elevation in adversarial training of the 
perceptual decoder, the distortion loss is also used in combination 
with the adversarial loss to achieve a satisfactory performance in practice.

\section{Experiments}
\subsection{Datasets and Training}
We train our model on various noise conditions using clean and noisy speech datasets.
For clean speech, we use the train-clean-100 and train-clean-360 of LibriTTS \cite{libritts}
as clean speech dataset, which has a sampling rate of 24 kHz. LibriTTS corpus is constructed
from audiobooks and includes the speech data from 2456 different speakers. In data pre-processing,
we filter out the samples which are resampled from 16 kHz to 24 kHz in the LibriTTS, resulting in 147k clean samples.
We generate noisy speech by mixing natural noise with clean speech at a sampling rate of 24 kHz.
We use Freesound \cite{freesound} as natural environmental noise. Freesound contains music,
mechanical noise, natural atmosphere and many other kinds of audio. We choose unlicensed audio
without human voice as noise data. Different noise segments are inserted into clean speech every
three seconds. The signal-to-noise ratio (SNR) is uniformly distributed between 0 dB and 15 dB.
Finally, we train our model on 147k clean speech and 33k noisy speech.

To enable bit-rate scalability without retraining the model, Residual Vector Quantizer\cite{rvq}
and Quantizer Dropout\cite{soundstream} are used in the training of all our model. We cascade 24
vector quantizers, each with a codebook size of 1024, and then quantize the residuals iteratively.
Under this setting, our model can support up to 18 kbps target rate. By dropping out different numbers
of cascaded VQs in training, our model can support a target rate between 3 kbps and 18 kbps.

We train our model SEStream with Adam for 500 epochs, with a learning rate of \(10^{-4}\) and batch size of 64.
For fair comparison, we train SoundStream and SEStream with the same dataset, the same network structure,
and the same hyper parameters in the loss function. The denoising flag of SoundStream is turned on 50\%
of the time during training. The input of the model is a randomly sampled 360 millisecond audio waveform.
The ground truth is the corresponding clean speech. The peak of each fragment is normalized to 0.95 and
multiplied by a gain between 0.3 and 1.0.

\subsection{Evaluation Metrics}
In this work, we use Virtual Speech Quality Objective Listener (ViSQOL) \cite{visqol}
as the evaluation metric for objective quality. It uses the similarity of the time-frequency
spectrum between the reference audio and the test audio to model the perceptual quality of
human speech, which is highly correlated with distortion loss. In addition, we use other
objective evaluation metrics, including Short-Time Objective Intelligibility (STOI) \cite{STOI},
CSIG, CBAK and COVL, to evaluate the quality of reconstructed audio. STOI is used to evaluate
the intelligibility of noisy speech, which is masked in time domain or weighted in frequency domain
after short time Fourier transform. CSIG predicts the mean opinion score (MOS) of speech distortion,
CBAK predicts the MOS of background noise, whilst COVL predicts the MOS of overall processed speech quality.
STOI ranges from 0.0 to 1.0 and other metrics range from 0.0 to 5.0. For all these evaluation metrics,
higher value represents better performance.

For subjective quality, we use Multi-Stimulus Test with Hidden Reference and Anchor (MUSHRA) \cite{mushra}
to compare our model and other representative audio codecs. MUSHRA is a double-blind listening test, in which test listeners rate the relative quality of the output of each codec against a hidden reference.
The reference audio is clean speech, together with decoded results by different codecs in the
presence of various background noise, are rated by the listeners.
The scores are ranged from 0 to 100 and a higher score reflects a higher subjective quality.

\section{Results}

\subsection{Objective  Quality}
\begin{table*}
\centering
\label{tab:objective}
\caption{Comparison on objective evaluation metrics at a target bit-rate of 6 kbps under different noise levels with quantizer dropout.}
\resizebox{\textwidth}{!}
{\begin{tabular}{ccccccccccc}
\hline
\multirow{2}{*}{\makecell[c]{Input \\ SNR}} & \multicolumn{2}{c}{ViSQOL} & \multicolumn{2}{c}{STOI} & \multicolumn{2}{c}{CSIG} & \multicolumn{2}{c}{CBAK} & \multicolumn{2}{c}{COVL} \\
                           & SoundStream       & SEStream & SoundStream       & SEStream   & SoundStream      & SEStream     & SoundStream       & SEStream    & SoundStream       & SEStream      \\ \hline
0 dB                        & 3.45         & 3.48        & 0.62        & 0.63       & 2.11        & 2.21       & 1.60        & 1.61       & 1.51        & 1.57       \\
5 dB                        & 3.70         & 3.74        & 0.73        & 0.74       & 2.59        & 2.65       & 1.83        & 1.82       & 1.86        & 1.90       \\
10 dB                       & 3.84         & 3.90        & 0.78        & 0.78       & 2.89        & 2.91       & 1.97        & 1.93       & 2.10        & 2.10       \\
15 dB                       & 3.96         & 4.01        & 0.80        & 0.79       & 3.06        & 3.05       & 2.05        & 1.99       & 2.24        & 2.22       \\
Clean                      & 4.05         & 4.12        & 0.80        & 0.80       & 3.18        & 3.15       & 2.10        & 2.02       & 2.33        & 2.28       \\ \hline
\end{tabular}}
\end{table*}

\begin{figure}
    \centering
    \includegraphics[width=0.48\textwidth]{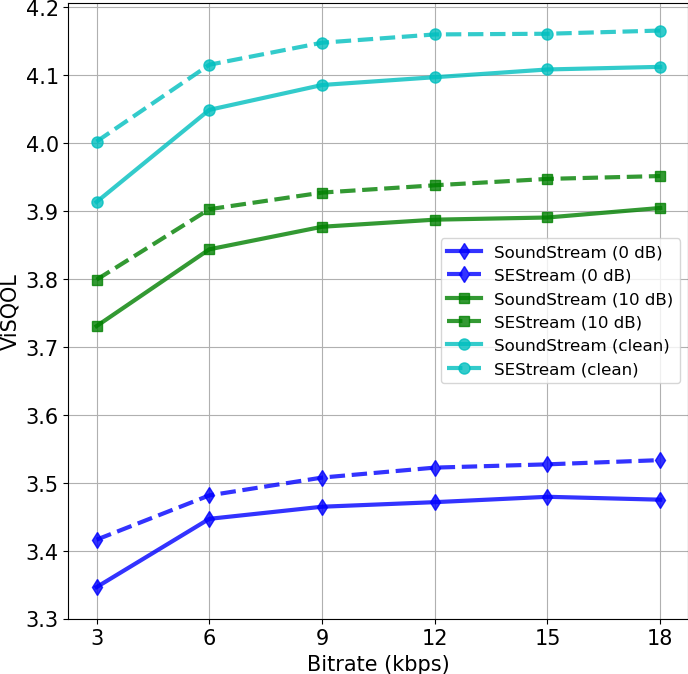}
    \caption{ViSQOL scores at different bit-rates (3 kbps $\sim$ 18 kbps) on clean test set and several noisy test set with different SNRs (0 dB, 10 dB). SEStream and SoundStream are compared under the same parameter setting and the same network structure.}
    \label{fig:visqol}
\end{figure}

In our experiment, we randomly select 200 clean samples from the
test set of LibriTTS as clean test set, which is disjoint with the training set.
To evaluate the denoising effect, natural noise with SNRs of 0 dB, 5 dB, 10 dB and 15 dB
were added to the 200 samples respectively to generate the noisy test set.
The synthesis criterion is the same as mentioned in Section IV-A.
Each audio in the test set lasts 2 to 4 seconds.

%In our experiment, we randomly selected 200 clean samples from the test set of LibriTTS as clean test set, which is disjoint with the training set. To evaluate the denoising effect, natural noises with SNRs of 0 dB, 5 dB, 10 dB and 15 dB were added to 200 samples respectively to generate the noisy test set. The synthesis criterion is the same as shown in section (IV-A). Each audio in the test set lasts 2 to 4 seconds.

To evaluate the objective quality, 
we compare the proposed SEStream with SoundStream 
in terms of ViSQOL test. Fig. \ref{fig:visqol} shows 
the ViSQOL scores evaluated on clean test set 
and several noise test sets with different SNRs. 
For SoundStream, the denoising flag is on for the
conditions of noisy speech to realize denoising.
By controlling the number of cascaded VQs during 
inference, we show the objective quality scores 
at target bit-rates from 3 kbps to 18 kbps.
As shown in Figure~\ref{fig:visqol}, the proposed SEStream
can achieve consistently better ViSQOL score than SoundStream
at all the tested bit-rates,
increasing the ViSQOL score by up to 0.089 (at 3 kbps bit-rate and clean set) and with an average improvement of about 0.057 on the
test cases.

%is always better than SoundStream at any bit-rate, increasing the ViSQOL score by up to 0.089 with an average increase of 0.057. This shows that the optimal training framework can obtain less distortion under the same conditions.

%Keep the denoising flag of SoundStream on during the inference to realize denoising. By controlling the number of cascaded VQs during inference, we show the objective quality scores at target bit-rates from 3 kbps to 18 kbps. As shown in Figure~\ref{fig:visqol}, our framework is always better than SoundStream at any bit-rate, increasing the ViSQOL score by up to 0.089 with an average increase of 0.057. This shows that the optimal training framework can obtain less distortion under the same conditions.

To further compare the denoising performance of the two methods, 
we select the target bit-rate of 6 kbps and use STOI, CSIG, CBAK, 
and COVL to evaluate the denoising quality under different noise levels.
The result is shown in Table I. It can be seen that in the case 
of relatively large background noise, especially when the SNR is 0 dB, 
our proposed method has better performance than SoundStream in most cases
in terms of the evaluation metrics. This indicates SEStream can 
generate speech with higher intelligibility and less background noise. 
In the cases of relatively low noise, e.g., with less background noise,
the advantage of SEStream vanishes in terms of the STOI, CSIG, CBAK, and COVL scores, but SEStream still achieves better ViSQOL score.

Fig. \ref{fig:spectrum} compares the time-frequency spectrum of 
two examples generated by SoundStream and SEStream under 0 dB SNR and 10 dB SNR. Both SoundStream and SEStream can reconstruct the speech from 
noisy speech that contains severe environmental noise. 
By comparing the spectrum of each sample longitudinally, 
it can be seen that SEStream can suppress the background noise 
better than SoundStream, as shown in the blue boxes. 
In the green boxes, SEStream retains more information of 
the clean spectrum. This phenomenon is relatively obvious 
in the case of severe noise.

\subsection{Comparison with Other Codecs}

%We generate 20 examples for low bit-rate and 20 examples for medium bit-rate, which contain clean test set and noisy test set with SNRs ranging from 0 dB to 15 dB uniformly. To further compare the subjective quality at low bit-rate and severe noise, we generate another 15 examples with 0 dB SNR and 15 examples with 10 dB SNR. The final result guarantees that each audio is rated more than 10 times. Audio examples are provided publicly online\footnote{\textit{https://jscscloris.github.io/SEStream/demo-audio/demo.html}}.

\begin{figure*}[!ht]
    \centering
    \subfigure[Low target bit-rate.]{
    \includegraphics[width=0.44\textwidth]{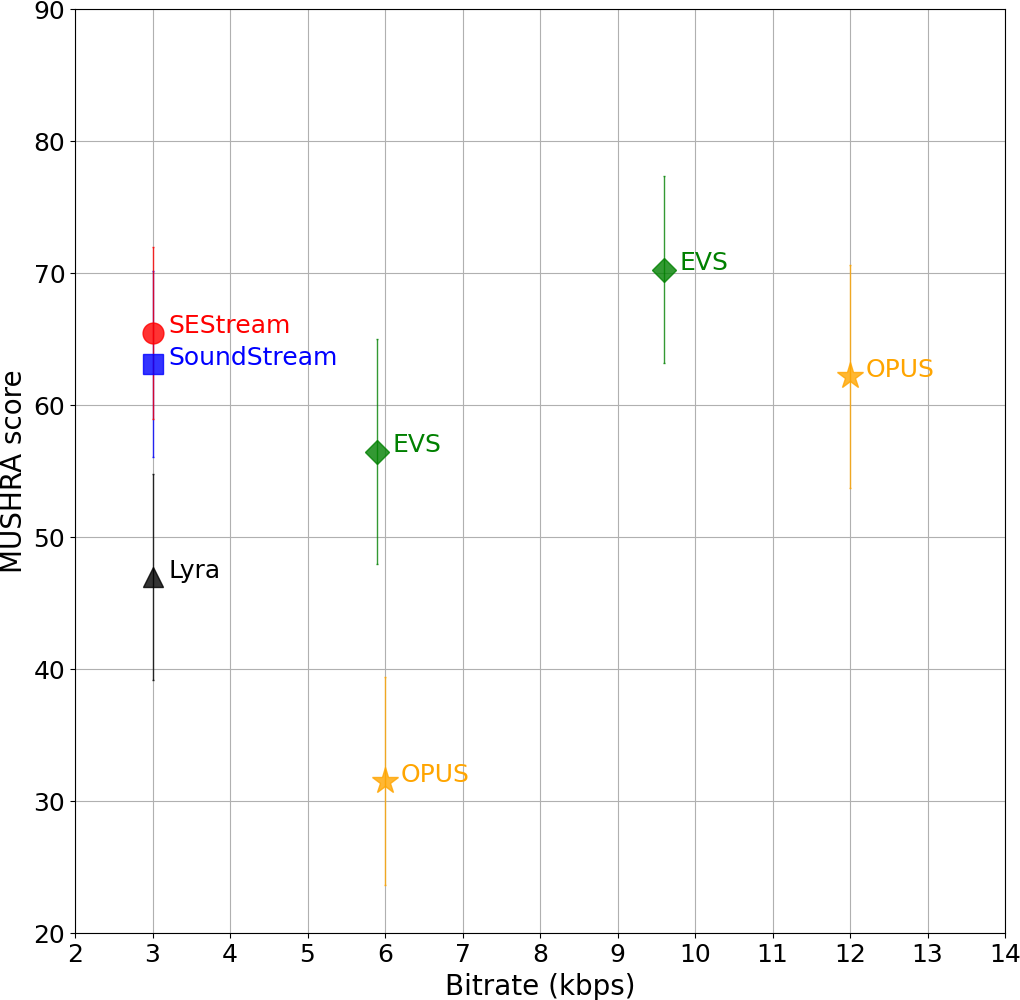}
    }
    \hspace{0.5in}
    \subfigure[Medium target bit-rate.]{
    \includegraphics[width=0.44\textwidth]{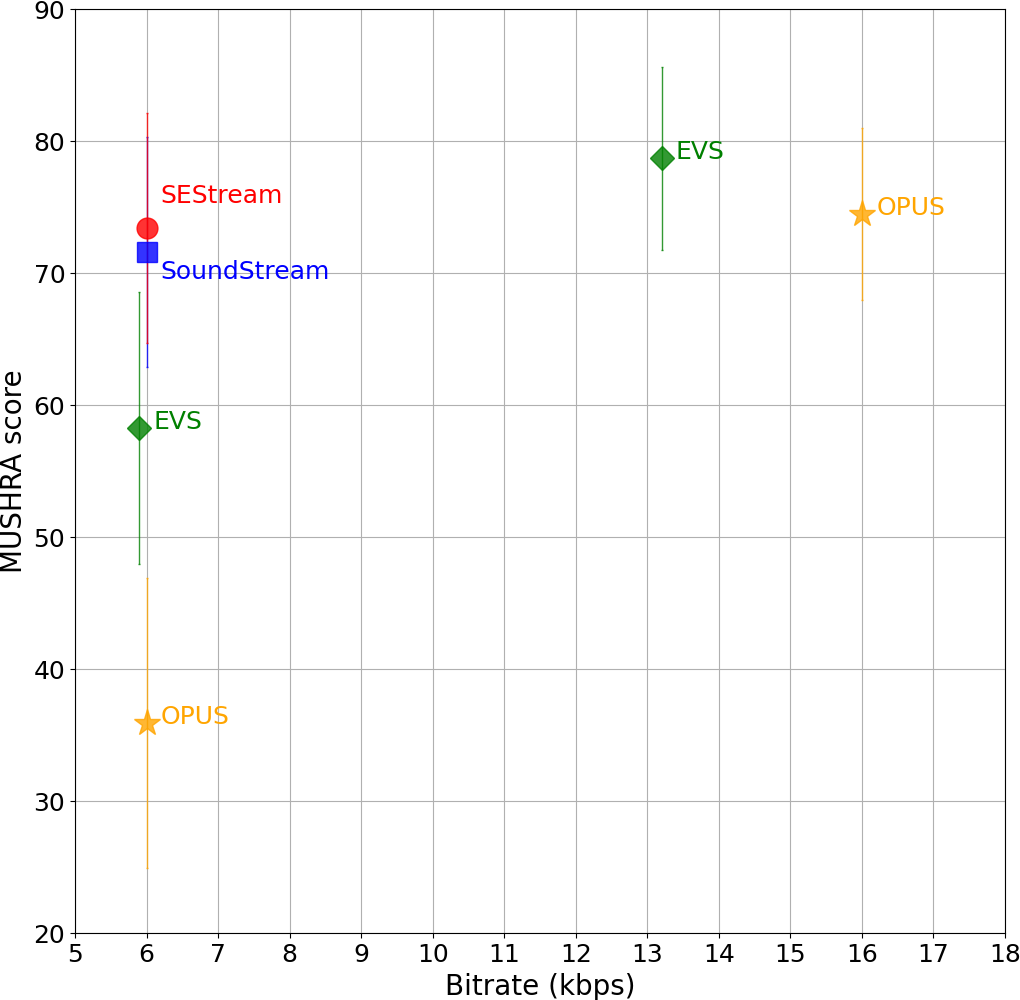}
    }
    \caption{Comparison of subjective quality scores of different codecs based on MUSHRA scores. Error bars denote 95\% confidence intervals.}
    \label{fig:mushra}
\end{figure*}
\begin{figure*}[!ht]
    \centering
    \subfigure[Low target bit-rate.]{
    \includegraphics[width=0.44\textwidth]{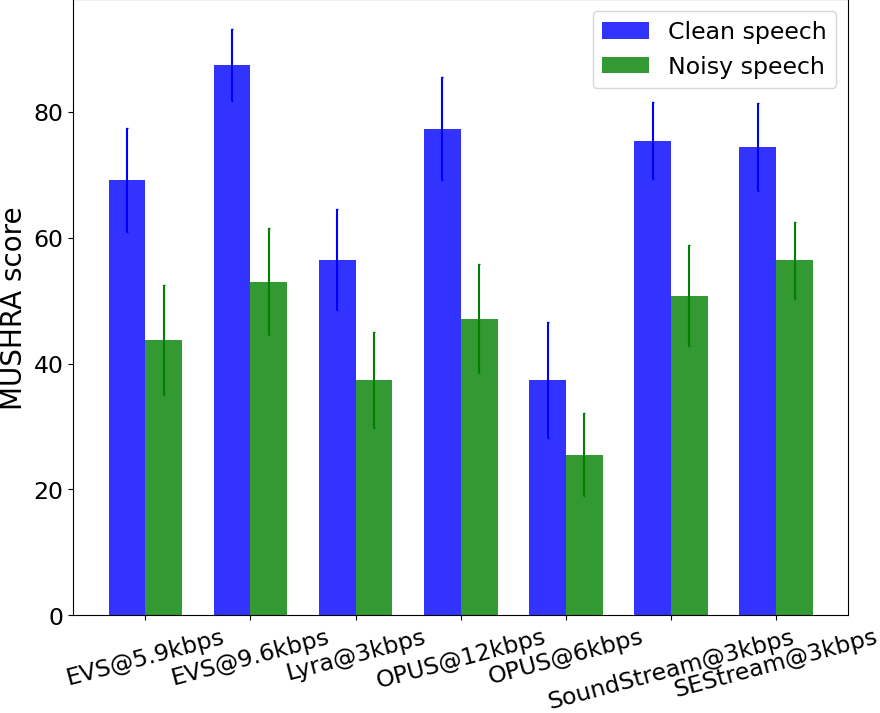}
    }
    \hspace{0.5in}
    \subfigure[Medium target bit-rate.]{
    \includegraphics[width=0.44\textwidth]{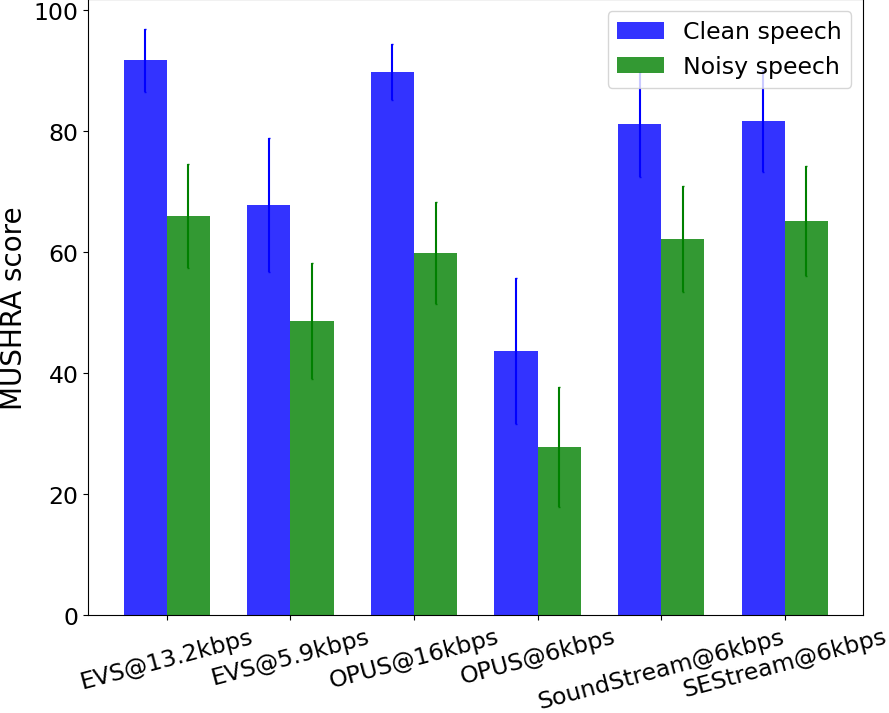}
    }
    \caption{Details of MUSHRA scores, which is divided into clean test set and noisy test set. Error bars denote 95\% confidence intervals.}
    \label{fig:mushra_d}
\end{figure*}
\begin{figure*}[!ht]
    \centering
    \subfigure[SNR = 0 dB.]{
    \includegraphics[width=0.44\textwidth]{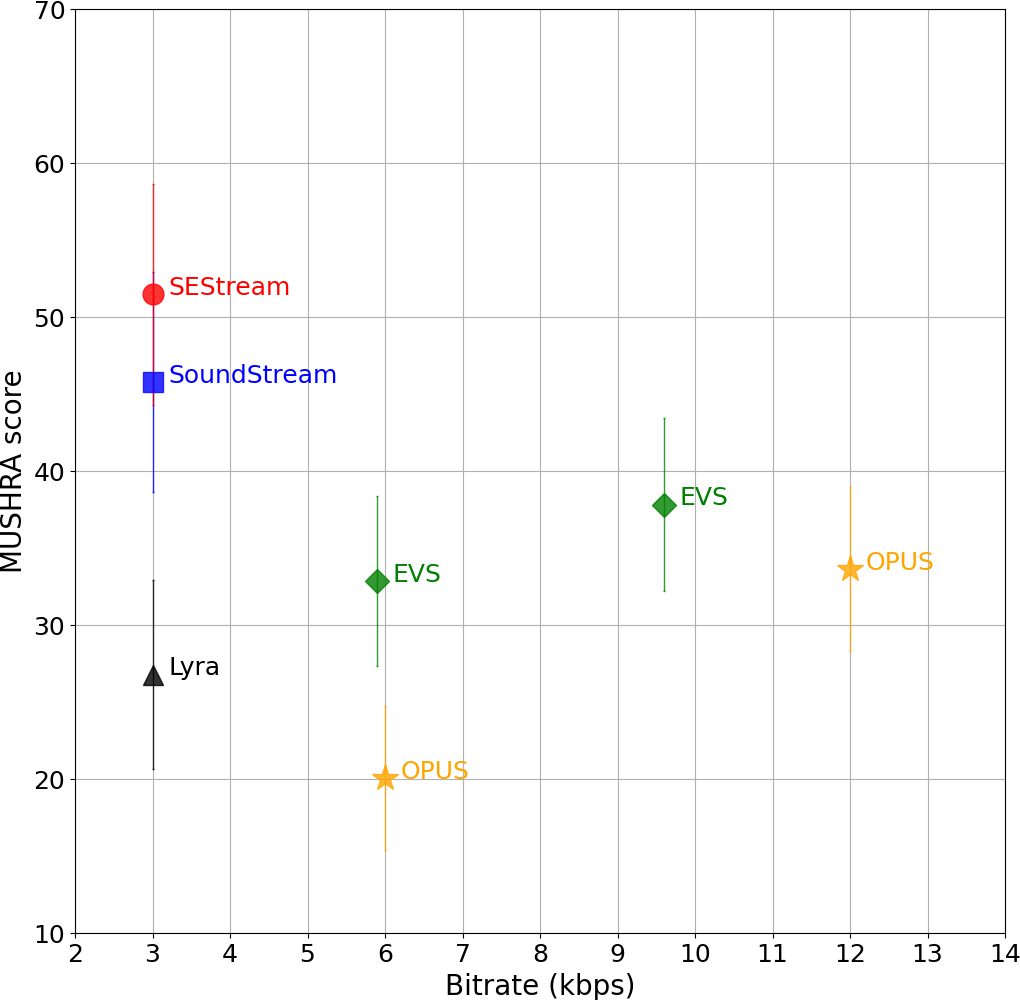}
    }
    \hspace{0.5in}
    \subfigure[SNR = 10 dB.]{
    \includegraphics[width=0.44\textwidth]{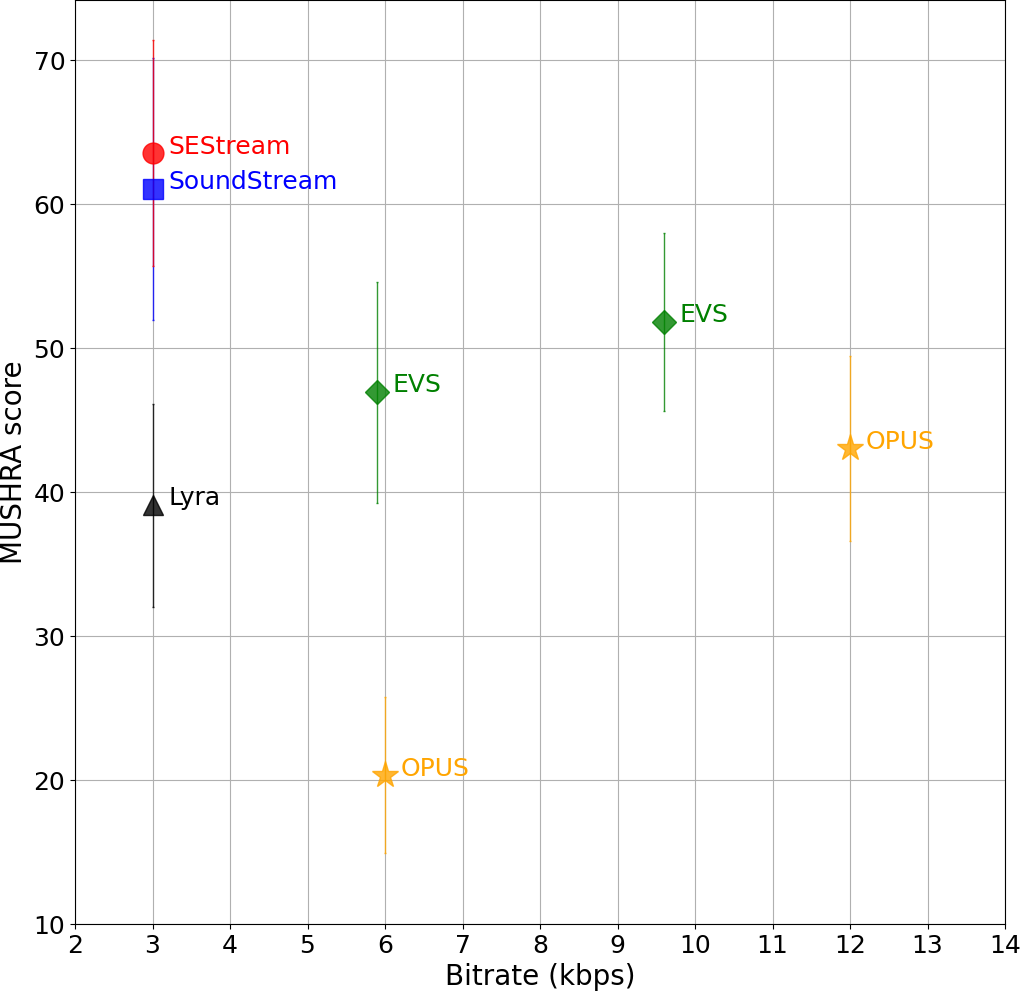}
    }
    \caption{Comparison of subjective quality scores of different codecs in low bit-rate and low SNR in terms of MUSHRA score. Error bars denote 95\% confidence intervals.}
    \label{fig:mushra_noisy}
\end{figure*}

We further compare our method with other representative audio codecs,
including Lyra, EVS, and OPUS at various bit-rates. 
We compare the methods on 20 examples for each case of 
low bit-rate and medium bit-rate, which contain clean test set and noisy test set with SNRs ranging from 0 dB to 15 dB uniformly.
The subjective quality score based on MUSHRA is shown in Fig. \ref{fig:mushra}. Details are shown in Fig. \ref{fig:mushra_d}.
It can be seen that at low and medium bit-rates, 
our method achieves a performance comparable with SoundStream 
in terms of subjective quality score with a slight advantage. 
Using a low bit-rate, e.g., 3 kbps, SoundStream and SEStream 
can achieve scores close to that of EVS at a higher bit-rate, e.g., 9.6 kbps, while outperforming OPUS at a higher bit-rate of 12 kbps.

%For noisy speech, the advantage is more conspicuous,
%reflecting the better performance of SoundStream and SEStream
%in the presence of noise.
%In the case of low bit-rate, all the methods based on the SoundStream backbone can approach the results close to EVS at a higher bit-rate. 
%, while our method successfully integrates noise reduction into the end-to-end codec process. 

As can be seen from Fig. \ref{fig:mushra_d}, the perceptual quality 
of our proposed method on clean speech is similar to that of SoundStream,
and the advantage is in noisy speech at a low bit-rate. 
Compared with other codecs, it is obvious that the joint speech 
compression and enhancement has better perceptual quality on noisy speech.
%For noisy speech, the advantage of SoundStream and SEStream over other
%codecs is more conspicuous, reflecting the better performance of SoundStream and SEStream in the presence of noise.

Fig. \ref{fig:mushra_noisy} compares the subjective quality score 
in the cases of low bit-rate and high noise with 0 dB and 10 dB SNR.
In these cases, the performance of traditional codecs deteriorates
drastically. In comparison, the advantage of SoundStream and SEStream 
over other codecs is more conspicuous, reflecting the better performance 
of SoundStream and SEStream in the presence of noise.

%can hardly filter out this serious noise and even lead to serious speech distortion. Compared with SoundStream, TTStream can filter out more noise and retain effective information at once.

Table II compares the codecs in terms of the objective metric ViSQOL.
We select 200 examples for the clean test set and 200 examples for the noisy test set (with SNRs ranging from 0 dB to 15 dB). SEStream has better ViSQOL scores than SoundStream, indicating that the distortion of SEStream is lower. Although the denoised results of Lyra sound relatively natural, there exists more serious distortion in timbre, thus, with significantly 
worse ViSQOL scores.

%Since human is not sensitive to the distortion of speech, we compare our results with other traditional codecs using the objective metric ViSQOL as shown in Table~\ref{tab:othercodec}. We selected 200 examples for the clean test set and 200 examples for the noisy test set (SNRs range from 0 dB to 15 dB). SEStream scored slightly higher on ViSQOL than SoundStream, suggesting that the distortion in SEStream is smaller. However, human is not sensitive to the advantage of distortion. Although the denoised results of Lyra sound relatively natural, there exists very serious distortion in timbre.

\begin{table}
\centering
\caption{Comparison of objective quality of different codecs in terms of ViSQOL score.}
\label{tab:othercodec}
\begin{tabular}{lll}
\hline
             & Clean & Noisy \\ \hline
Lyra @3 kbps   & 2.43  & 2.37  \\
EVS @5.9 kbps  & 2.46  & 2.40  \\
EVS @9.6 kbps  & 3.90  & 3.50  \\
EVS @13.2 kbps & 3.94  & 3.61  \\
OPUS @6 kbps   & 2.13  & 2.07  \\
OPUS @12 kbps  & 2.68  & 2.56  \\
OPUS @16 kbps  & 4.11  & 3.64  \\ \hline
SoundStream @3 kbps & 3.91  & 3.66  \\
SoundStream @6 kbps & 4.05  & 3.76  \\
SEStream @3 kbps  & 4.00  & 3.71  \\
SEStream @6 kbps  & 4.12  & 3.82  \\ \hline
\end{tabular}
\end{table}

\begin{figure*}[!ht]
    \centering
    %first row
    \subfigure{
    \rotatebox{90}{\scriptsize{Noisy speech}}
    \begin{minipage}[t]{0.23\linewidth}
      \centering
      \includegraphics[width=1\linewidth]{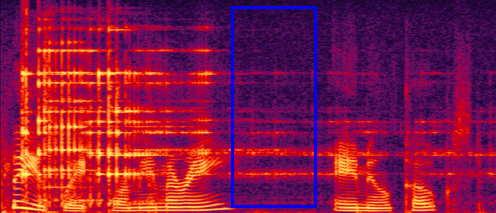}
    \end{minipage}
    \begin{minipage}[t]{0.23\linewidth}
      \centering
      \includegraphics[width=1\linewidth]{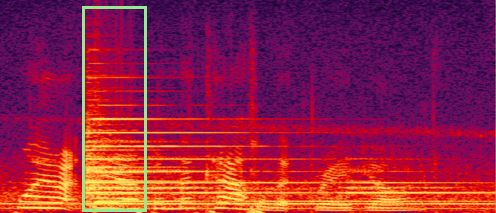}
    \end{minipage}
    }
    \subfigure{
    \begin{minipage}[t]{0.23\linewidth}
      \centering
      \includegraphics[width=1\linewidth]{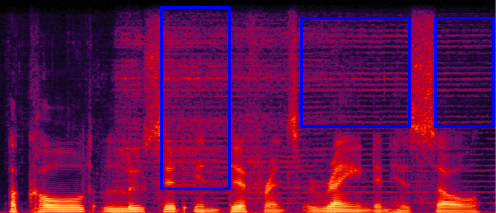}
    \end{minipage}
    \begin{minipage}[t]{0.23\linewidth}
      \centering
      \includegraphics[width=1\linewidth]{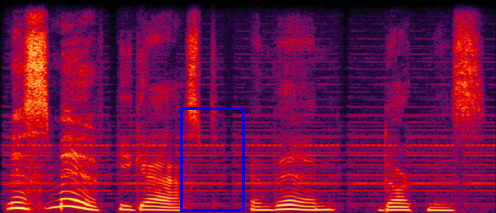}
    \end{minipage}
    }
    \setcounter{subfigure}{0}
    %second row
    \subfigure{
    \rotatebox{90}{\scriptsize{Clean speech}}
    \begin{minipage}[t]{0.23\linewidth}
      \centering
      \includegraphics[width=1\linewidth]{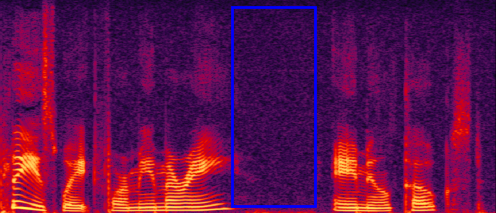}
    \end{minipage}
    \begin{minipage}[t]{0.23\linewidth}
      \centering
      \includegraphics[width=1\linewidth]{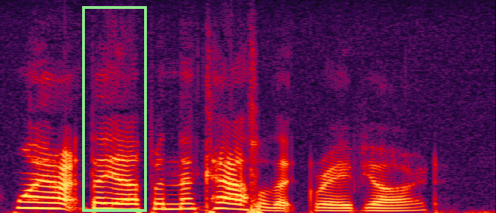}
    \end{minipage}
    }
    \subfigure{
    \begin{minipage}[t]{0.23\linewidth}
      \centering
      \includegraphics[width=1\linewidth]{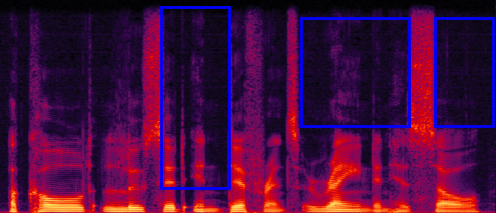}
    \end{minipage}
    \begin{minipage}[t]{0.23\linewidth}
      \centering
      \includegraphics[width=1\linewidth]{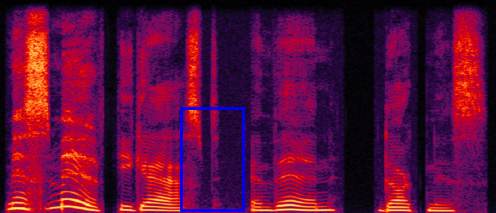}
    \end{minipage}
    }
    \setcounter{subfigure}{0}
    %third row
    \subfigure{
    \rotatebox{90}{\scriptsize{SoundStream}}
    \begin{minipage}[t]{0.23\linewidth}
      \centering
      \includegraphics[width=1\linewidth]{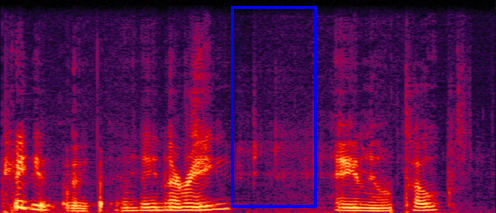}
    \end{minipage}
    \begin{minipage}[t]{0.23\linewidth}
      \centering
      \includegraphics[width=1\linewidth]{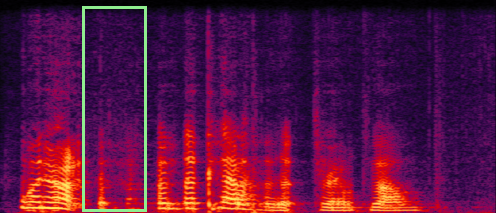}
    \end{minipage}
    }
    \subfigure{
    \begin{minipage}[t]{0.23\linewidth}
      \centering
      \includegraphics[width=1\linewidth]{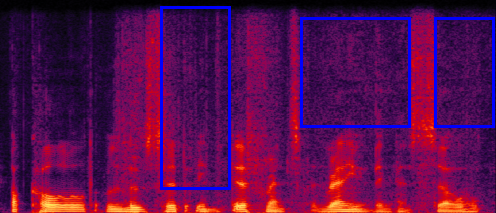}
    \end{minipage}
    \begin{minipage}[t]{0.23\linewidth}
      \centering
      \includegraphics[width=1\linewidth]{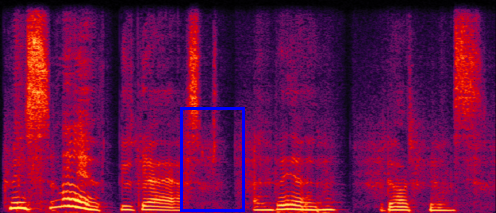}
    \end{minipage}
    }
    \setcounter{subfigure}{0}
    %forth row
    \subfigure[SNR = 0 dB.]{
    \rotatebox{90}{\scriptsize{SEStream}}
    \begin{minipage}[t]{0.23\linewidth}
      \centering
      \includegraphics[width=1\linewidth]{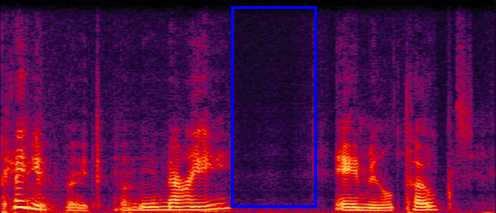}
    \end{minipage}
    \begin{minipage}[t]{0.23\linewidth}
      \centering
      \includegraphics[width=1\linewidth]{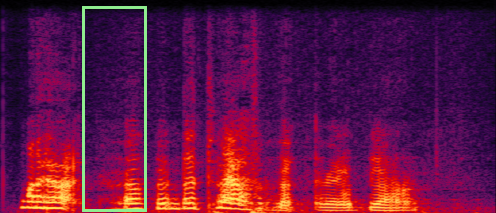}
    \end{minipage}
    }
    \subfigure[SNR = 10 dB.]{
    \begin{minipage}[t]{0.23\linewidth}
      \centering
      \includegraphics[width=1\linewidth]{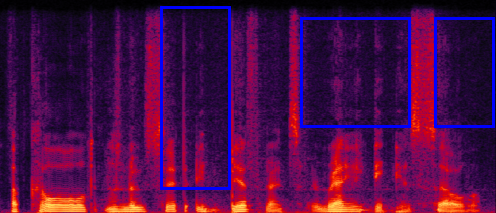}
    \end{minipage}
    \begin{minipage}[t]{0.23\linewidth}
      \centering
      \includegraphics[width=1\linewidth]{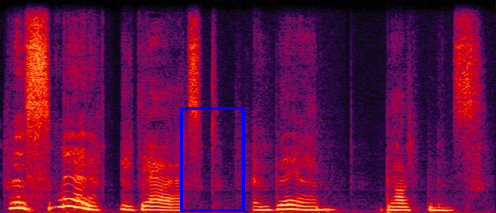}
    \end{minipage}  
    }    
    \caption{Qualitative comparison on the results of SoundStream and the proposed SEStream in two noise conditions with SNR = 0 dB and SNR = 10 dB, respectively.}
    \label{fig:spectrum}
\end{figure*}

\section{Conclusions}

In this paper, a theoretical result is derived for
the joint compression and enhancement problem for speech 
signal in the presence of background noise. Based on the
result, a two-stage training method for joint speech
compression and enhancement has been proposed. 
Extensive experimental results on various bit-rate and noise
conditions demonstrated that, the 
proposed method can achieve better performance in comparison
with state-of-the-art codecs in terms of both objective and
subjective metrics. The advantage of the proposed method
is more distinct in the case of low bit-rate and high noise.

%propose an optimal training framework based on the backbone of SoundStream. The core contribution of our paper is that extend lossy compression theory to noise and generalized distortion situations. We use the formula to prove theoretically that the proposed framework is still optimal under joint compression and enhancement in audio. We train the SoundStream and our proposed framework with the same parameter and test under different SNR conditions.  Experiments show that our method has advantages in both objective and subjective evaluation over raw SoundStream.

%\section{Acknowledgements}
%This work was supported partially by the STI 2030-Major Projects (Grant no. 2022ZD0208700), the Shanghai Municipal Science and Technology Major Project (Grant no. 2021SHZDZX0102) and MoE Key Lab of Artificial Intelligence, AI Institute, Shanghai Jiao Tong University, China.

\bibliographystyle{IEEEtran}
\bibliography{sample}

\end{document}